\documentclass[journal]{IEEEtran}
\usepackage[dvipsnames]{xcolor}
\usepackage{cite}
\usepackage{amsmath,amssymb,amsfonts}
\usepackage{mathtools}
\usepackage{lipsum}
\usepackage{algorithmic}
\usepackage{graphicx}
\usepackage{textcomp}
\usepackage{booktabs}
\usepackage{esint}

\ifCLASSINFOpdf
\else
\fi
\graphicspath{{Figures/}}

\begin{document}
%
\title{Equivalent Circuit Models for Waveguide-Fed, Resonant, Metamaterial Elements}
%
%
%

\author{David R. Smith,~\IEEEmembership{Senior Member,~IEEE,}
        ~Yeonghoon Noh,~\IEEEmembership{Member,~IEEE,} Insang Yoo,~\IEEEmembership{Member,~IEEE,} Divya Pande and Mohammad Ranjbar Nikkhah %
\thanks{D. R. Smith is with the Department
of Electrical and Computer Engineering and the Center for Metamaterials and Integrated Plasmonics, Duke University,
Durham, NC, 27708 USA. I. Yoo is with the School of Electrical and Electronic Engineering, Yonsei University, Seoul, Korea. He was previously with the Department of Electrical and Computer Engineering and the Center for Metamaterials and Integrated Plasmonics, Duke University, Durham, NC, 27708 USA. e-mail: insang.yoo@yonsei.ac.kr. M. Ranjbar Nikkhah is with Kymeta Corp., Redmond, WA, 98052, USA}
\thanks{Manuscript received ---, 2024; revised ---, 2024.}}

%
%

\markboth{IEEE Transactions on Antennas and Propagation,~Vol.~X, No.~X, ~2024}%
{Shell \MakeLowercase{\textit{et al.}}: Bare Demo of IEEEtran.cls for IEEE Journals}
%



\maketitle

\begin{abstract}
We propose an approach to extracting equivalent circuit models for waveguide-fed, resonant metamaterial elements, such as the complementary, electric inductive-capacitive element (cELC). From the scattering parameters of a single waveguide-fed cELC, effective electric and magnetic polarizabilities can be determined that can be expressed in terms of equivalent lumped element circuit components. The circuit model provides considerable insight into the electromagnetic scattering properties of cELCs as a function of their geometric parameters and imparts intuition useful for element optimization. We find that planar, inherently resonant, waveguide-fed elements exhibit a set of common properties that place constraints on their coupling, maximum radiation, and other key scattering parameters. In addition, unlike simple slots and other non-resonant irises, resonant elements introduce an effective transformer to the equivalent circuit that accounts for the field enhancement occurring in such elements at resonance. We introduce a general and robust method to determine the effective circuit parameters by fitting to the extracted polarizability, extending the approach to resonant metamaterial elements integrated with physical lumped circuit components, such as packaged capacitors or varactors. We find excellent agreement between the analytical predictions and full-wave simulations, such that with one or two full-wave simulations the properties of the cELC can be determined for any externally added lumped elements. This approach can be leveraged to dramatically increase the efficiency of metasurface aperture design, especially when libraries of element responses are required.
\end{abstract}


%
\IEEEpeerreviewmaketitle

\section{Introduction}
%
%
%
%
\IEEEPARstart{M}{etasurfaces}---collections of scattering elements distributed over a surface---represent one of the most successful outcomes of the metamaterials field in terms of practicality \cite{holloway2012overview, wang2020metantenna, minatti2014modulated, maci2011metasurfing, quevedo2019roadmap, glybovski2016metasurfaces, li2018metasurfaces, chen2016review, yu2014flat}. Metasurface variants have been used across the electromagnetic spectrum, from radio frequency (RF) to optical, for diverse applications including antennas, lenses, reflectors, and generalized quasi-optical components. Unlike volumetric metamaterials, metasurfaces have achieved widespread use and even commercialization because they provide unique and tailored functionality with minimal losses. Moreover, metasurfaces can be readily combined with conventional radio frequency (RF) and optical devices, such as waveguide feeds and sources, enabling metasurfaces to replace or augment components in otherwise conventional systems.

Metamaterial elements---especially resonators that are based on conducting materials---inherently possess losses that can lead to significant absorption of waves. Metasurfaces effectively reduce the number of elements encountered by the incident wave, so that overall losses can be mitigated and competitive devices constructed. Numerous variants of metasurface apertures have been proposed and demonstrated, including reflectarrays (such as reconfigurable intelligent surfaces, or RIS) \cite{campbell2020reflectarray, elmossallamy2020ris, farmahini2013metasurface} and surface wave architectures \cite{sievenpiper2003two, maci2011metasurfing, patel2013effective}, which consist of relatively simple metamaterial elements tiled over a surface to produce or steer beams, or shape desired radiation patterns. Common to these platforms are passive, subwavelength, scattering elements whose scattering properties dictate the range of phase and amplitude control available at each point throughout the composite aperture. Whenever control over the distribution of the amplitude and phase of the aperture field is available, arbitrary radiation patterns can be produced.

\begin{figure}[!b]
\centering
\includegraphics[width=2.75in]{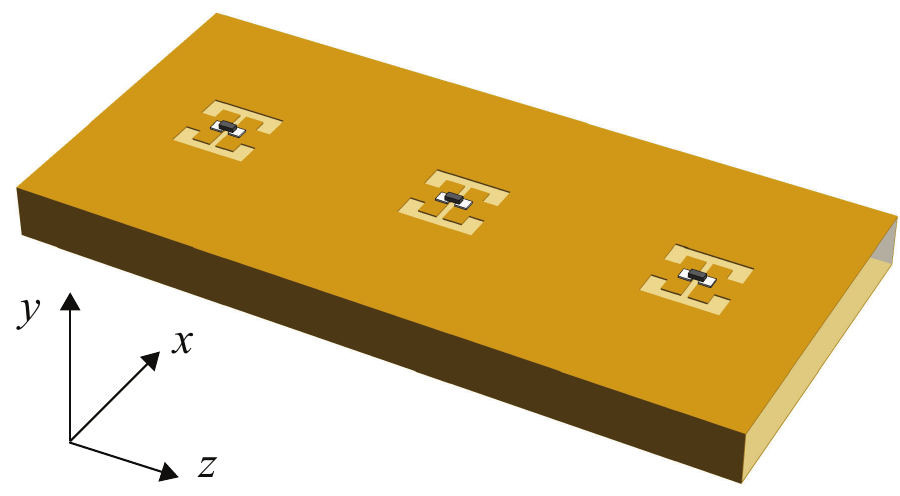}
\caption{A rectangular waveguide coupled to several cELC metamaterial elements, each with an integrated, packaged capacitor. The dimensions of the elements here have been exaggerated for clarity.}
\label{fig:illustration_capacitor_loaded_celcs}
\end{figure}

The particular variant of metasurface of interest here is the waveguide-fed metasurface, illustrated in Fig. \ref{fig:illustration_capacitor_loaded_celcs}, which consists of a waveguide feed structure that couples to an array of metamaterial elements \cite{smith2017analysis, li2020widebandmetasurface, johnson2014discrete}. Waveguide-fed metasurfaces that can be dynamically tuned---for example, with the inclusion of electronically controlled elements or materials---have gained considerable traction over the past decade as alternatives to phased arrays and electronically scanned antennas (ESAs) \cite{johnson2014discrete,johnson2016extremum,sleasman2017experimental, boyarsky2021electronically}. While waveguide-fed metasurfaces have transitioned to commercial products for a diverse set of applications, all variants of metasurfaces have similar potential, motivating the continued investigation of their fundamental operation. 

Inherent to the metamaterial concept is that each metamaterial element is significantly smaller than the excitation wavelengths. For waveguide-fed metasurfaces, the metamaterial element is often an iris that couples energy from the waveguide region to the free-space region. Irises significantly smaller than the operating wavelength can be described to good approximation as a combination of polarizable electric and magnetic dipoles---a description that dates back decades to work by Bethe and others \cite{bethe1944theory}. We refer to this description here as the \emph{dipole framework}. Though approximate, the conceptual substitution of small apertures with equivalent dipoles nevertheless provides considerable insight to the scattering problem and is often quite accurate when used in waveguide coupling problems or certain types of aperture antennas, such as slot arrays \cite{oliner1957impedance, stevenson1948waveguideslots}. The dipole framework extends the single aperture model to large composite apertures comprising hundreds or even thousands of metamaterial elements; since the dipole properties of each metamaterial element can be found from a full-wave simulation over a relatively small computational domain, the composite aperture can be modeled as a collection of polarizable dipoles. The far-fields radiated by the aperture, for example, are obtained by summing over the contributions of each dipole---a much faster and simpler process than would be required from a full-wave simulation of the entire aperture. The viability of this approach as a computational tool depends on the ability to extract accurate effective polarizability values for an arbitrary metamaterial element.

The extraction of effective polarizability values from full-wave simulations on arbitrary metamaterial elements embedded in waveguide structures has been previously developed and is largely straightforward \cite{pulido2017discrete,pulido2017polarizability}. From the measured or simulated scattering (S-) parameters corresponding to a single metamaterial element (or iris) embedded in one of the conducting walls of a waveguide, the effective polarizability values can be extracted and subsequently used for modeling the composite aperture. However, for each metamaterial element geometry, a full-wave simulation must be performed to determine the polarizabilities; similarly, if lumped components are added to the design, full-wave simulations must be performed to determine the resultant properties. While it is possible to extract the polarizability for any given element, this process does not provide much insight as to the origins of the polarizability that would assist in design revisions and optimization. In particular, since added lumped elements have known circuit models, it should be possible to understand immediately their impact on the scattering properties of an element if the polarizability model of that element equates to a circuit model.  

A recent analysis has established the equivalence of an effective circuit model with the polarizability model for slots and slot-fed patch elements, two elements that have been used in waveguide-fed metasurface apertures \cite{smith2024circuit}. Another type of common metamaterial element---and our focus for this present work---is the complementary electric inductive/capacitive (cELC) resonator. The cELC is the electromagnetic complement of the ELC, an element designed to exhibit a predominantly \emph{electric} response to incident electromagnetic fields \cite{schurig2006celc}. By interchanging the vacuum and conducting regions of the ELC, the Babinet equivalent of the ELC is formed, which has a predominantly \emph{magnetic} response to incident fields \cite{falcone2004babinet}. Commonly used in waveguide-fed metasurface designs, the cELC adds additional effective inductance and capacitance in an arrangement that motivates a somewhat more complicated equivalent circuit model than for simple rectangular slots. Being resonant even in the absence of additional lumped components, the cELC possesses interesting aspects beyond non-resonant metamaterial designs that must be correctly incorporated into the model.

In the present work, we suggest an equivalent circuit model for the cELC and demonstrate its validity through comparisons with full-wave simulations. As the metamaterial element increases in geometric complexity, there is the possibility for additional sources of parasitic circuit components that may not be immediately evident; nevertheless, we start with a simple, intuitive circuit model designed to capture what would likely be the major circuit contributions, and show that this model suffices to predict the scattering and radiating characteristics of the element. We begin with a lossless system, then show the impact of introducing either resistive or dielectric losses to the cELC element. We also demonstrate that the effective quality factor of the element can be predictably controlled by modifying the effective circuit parameters, which are in turn related to the geometry of the element.

Unlike simple coupling holes and irises that have been studied in prior works, the cELC has an intrinsic resonance related to its geometry that introduces a novel aspect to the coupling problem. Although the particular aspects of the cELC are somewhat unique, there are interesting parallels to general aperture coupling between waveguides and resonant cavities that provide useful guidance \cite{wheeler1964, collin1990fieldCh7, james1977}. Here, we use a network formulation developed for aperture coupling problems to justify the equivalent admittance of various cELC element designs \cite{harrington1976, harrington1991,liang1983generalized}. We find that the resonance of the cELC can be understood by the addition of an inductance and a capacitance that relate to the internal arms and the interior gap, respectively. In addition, a coupling factor must be included that introduces a transformer element into the cELC equivalent circuit; we determine that the transformer ratio is approximately related to the square root of the quality factor of the element's resonance. This factor is needed to account for the observed resonance frequency shift of the element when a lumped capacitance is inserted into the gap region.

In section \ref{sec:polarizabilityModel}, we introduce the lossless cELC polarizability and propose an equivalent circuit model. We then extend the model to cELC elements with resistive and dielectric losses. Using full-wave simulations, we demonstrate the validity of the polarizability extraction using several different cELC geometries, introducing an efficient method to determine the equivalent circuit parameters. Importantly, the simple equivalent circuit model proposed recovers the scattering behavior of the element exactly, with only three or four parameters required. The addition of a lumped capacitor is investigated in Section \ref{sec:celc_tuning}, where we introduce a method to assess the parasitic inductance of simulated lumped elements to arrive at an approximate model of a packaged capacitor. Applying the transformer correction as derived in the appendix, we show that the tuning of a cELC by the addition of a lumped capacitor can be predicted from the extracted circuit model.

\begin{figure}[!b]
\centering
\includegraphics[width=2.75in]{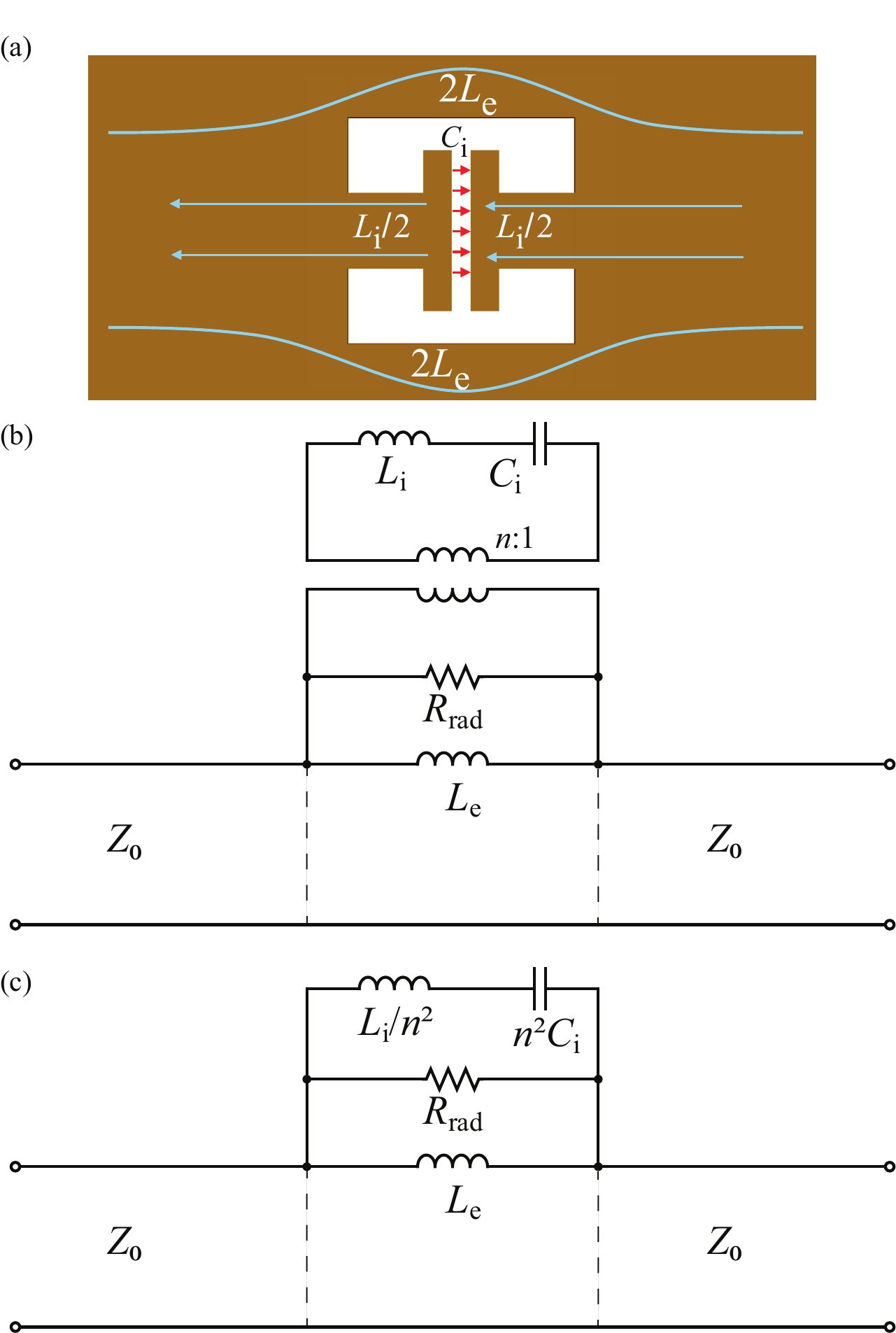}
\caption{(a) A waveguide-fed metamaterial element in the top conducting surface of a rectangular waveguide, with regions of inductance and capacitance indicated. (b) An equivalent circuit model with transformer coupling. (c) The same equivalent circuit model with transformed impedance.}
\label{fig:single_cELC}
\end{figure}

\section{cELC Polarizability and Circuit Models}
\label{sec:polarizabilityModel}

\subsection{Analytical Model}
A recent analysis has demonstrated the equivalence of circuit and polarizability models of slot and slot-fed patch elements within the context of a transmission line description \cite{smith2024circuit}. The polarizability description, inherent to the dipole framework and the theory of small apertures in conducting surfaces, assumes that the iris can be replaced by a set of point polarizable (electric and magnetic) dipoles located at the center of the element. The equivalent circuit model replaces these dipoles with a combination of series impedance and shunt admittance placed at the same point along a transmission line. In either description, the iris or metamaterial element must be signficantly subwavelength in dimension, such that the fields do not vary appreciably over the element. Furthermore, since only the magnetic dipole radiates to free space, it is desirable that the element be designed such that the magnetic dipole dominates the scattering, which equates to minimizing the shunt admittance in the equivalent circuit model. In \cite{smith2024circuit}, thin, rectangular slots in a waveguide oriented perpendicular to the direction of propagation were considered. A rectangular slot behaves as a polarizable dipole with an in-plane magnetic polarizability directed perpendicular to the propagation of the incident waveguide mode. Within the circuit model, the rectangular slot is equivalent to a series inductance inserted into a transmission line.

The specific metamaterial element we consider here is the complementary electric inductive-capacitive resonator (cELC), presented in Fig. \ref{fig:single_cELC}, which has been used frequently in waveguide-fed metasurface designs \cite{johnson2014discrete, yoo2016efficient, yoo2020full, boyarsky2021electronically, sleasman2016waveguide}. The effective iris that corresponds to the cELC is geometrically more complex than the simple rectangular slot previously considered. In a sense, the cELC can be thought of as a rectangular slot with parasitic inductance and capacitance spanning the slot, as indicated in Fig. \ref{fig:single_cELC}a. Here, we associate the centermost gap with a capacitance $C_i$ and the connecting arms with inductances $L_i/2$. These internal circuit elements are in parallel with the effective inductance of the overall slot, $L_e$, which relates to the currents deflected around the overall slot. Near resonance, it can be expected that the electric field energy will be concentrated mostly within the inner gap, leading to a field pattern similar to that of the rectangular slot. Thus, based on equivalence principles, the in-plane electric field can be related to an equivalent magnetic dipole \cite{pulido2017polarizability, collin1990field}, which will be the dominant component. 

A rigorous calculation of the scattered field distribution within and outside the waveguide produced by the cELC can be performed using the method-of-moments combined with equivalence principles and reciprocity \cite{harrington1991, collin1990field}. However, since the cELC is significantly subwavelength and we are interested in its properties primarily near resonance, we require only a few results from the theory to allow an equivalent circuit to be inferred. This brief analysis, provided in the appendix, motivates the circuit model shown in Fig. \ref{fig:single_cELC}b, where a transformer coupling has been introduced that connects the circuit to the internal capacitance and inductance. This equivalent circuit results from the effective coupling of the equivalent magnetic dipole moment to the cELC resonator.

The internal inductance and capacitance can be transformed so as to remove the transformer element from the equivalent circuit, as shown in Fig. \ref{fig:single_cELC}(c). If no lumped elements are added to the cELC, then the circuit models \ref{fig:single_cELC}(b) and \ref{fig:single_cELC}(c) provide the same information. To make the presentation simpler in this section, we use the circuit model of \ref{fig:single_cELC}c, with $L'_i=L_i/n^2$ and $C'_i=n^2 C_i$ being the effective internal inductance and capacitance.

The total impedance $Z$ for the model of \ref{fig:single_cELC}(c) can be expressed as

\begin{equation} \label{}
    Z=j\omega L_{e} \frac{1-\omega^2/\omega_1^2}{1-\omega^2/\omega_0^2},
    \label{eq:celc_impedance}
\end{equation}

\noindent where

\begin{equation}
    \omega_0^2 = \frac{1}{(L'_i+L_e)C'_i},
\end{equation}

\noindent and

\begin{equation}
    \omega_1^2 = \frac{1}{L'_i C'_i}.
\end{equation}

\noindent We assume here there are no resistive losses. Also, we do not include the radiation resistance term $R_{\textrm{rad}}$, which will be accounted for below using the exact analytical expressions.

As described above, the cELC produces predominantly a magnetic response so that we can ignore any shunt impedance and assume the cELC equivalent circuit appears as a series impedance in a transmission line. Under these assumptions, the effective magnetic polarizability of the cELC can be related to the equivalent impedance according to \cite{smith2024circuit}

\begin{equation}
    \alpha_m = \frac{\frac{-jabZ}{2\mu \omega}}{1+\frac{abZ}{2\mu \omega}\Big( \frac{\beta}{ab}+\frac{k^3}{3\pi}\Big)},
\end{equation}

\noindent an expression obtained by comparing the polarizability extraction from the waveguide scattering (S-) parameters with the transfer (or ABCD) matrix. The dynamic polarizability lies along the $\hat{x}$ direction and is complex due to the radiation damping terms; these terms can be obtained via energy conservation arguments or from the appropriate dipole Green's functions \cite{pulido2017polarizability, smith2024circuit}. Using the proposed impedance model for the cELC, given by (\ref{eq:celc_impedance}), the magnetic polarizability can be written as

\begin{equation}
    \alpha_m = \frac{\alpha_{m0} \omega_0^2 (1-\omega^2/\omega_1^2)}{\omega_0^2-\omega^2+j \alpha_{m0} \omega_0^2 (1-\omega^2/\omega_1^2)\Big( \frac{\beta}{ab}+\frac{k^3}{3\pi}\Big)},
    \label{eq:celc_polarizability}
\end{equation}

\noindent where

\begin{equation}
    \alpha_{m0}=\frac{a b L_e}{2 \mu},
\end{equation}

\noindent is the static polarizability---the value of the polarizability in the low frequency limit. The effective circuit parameters can be found by performing a fitting to either the numerically computed S-parameters of the waveguide-fed element or the extracted polarizability, as will be described in a later section. Note that for the lossless cELC, the polarizability is completely defined by three parameters: an amplitude, $\alpha_{m0}$, a pole, $\omega_0$, and a zero $\omega_1$.

In the absence of resistive losses, the quality factor of the resonance is set only by the radiation damping terms and is proportional to the static polarizability $\alpha_{m0}$. Since $\alpha_{m0}$ relates to the inductance of the element, the width of the resonance is proportional to the static inductance. The quality ($Q-$) factor can thus be modified by design of the element; keeping the resonance frequency fixed while changing the relative contributions of inductance and capacitance allows the Q-factor to be altered. This tunability may be of use in trading off instantaneous bandwidth for tuning bandwidth in reconfigurable metasurface designs.

It is useful to rewrite (\ref{eq:celc_polarizability}) in a somewhat different and more universal form. Since $\alpha_m$ is a complex function, we can find an expression for its argument as

\begin{equation}
    \tan{\theta} = -r\frac{\alpha_{m0}\omega_0^2(1-\omega^2/\omega_1^2)}{\omega_0^2-\omega^2}
    \label{eq:phase_of_resonance},
\end{equation}

\noindent where $r$ corresponds to the radiation damping terms,

\begin{equation}
    r=\Big( \frac{\beta}{a b}+\frac{k^3}{3\pi}\Big).
\end{equation}

\noindent Using the expression for the phase, (\ref{eq:phase_of_resonance}), in (\ref{eq:celc_polarizability}), we arrive at a general form for the resonance,

\begin{equation}
    \alpha_m = -\frac{1}{r(\omega)}\sin{\theta}e^{j\theta}.
\end{equation}

\noindent Although the radiation term $r(\omega)$ is dispersive, the majority of the phase change occurs at or near the resonance frequency of $\omega_0$. Thus, it is a reasonable approximation to treat $r$ as a constant, evaluated at $\omega_0$. The resulting polarizability is thus

\begin{equation}
    \alpha_m \approx -\frac{1}{r(\omega_0)}\sin{\theta}e^{j \theta}.
\end{equation}

\noindent The result of this analysis shows that \emph{the 
 polarizability for the cELC (as for any waveguide-fed element) has a maximum at resonance ($\theta=\pi/2$) that is determined entirely by the radiation damping terms}. No modification to the metamaterial element geometry will change the maximum polarizability, nor the fraction of power radiated versus scattered back into the waveguide. Changing the static polarizability, $\alpha_{m0}$, will change the apparent $Q$ of the resonance (\emph{i.e.}, the Q determined from the line width), but not its maximum value.

 Note that at the peak of the resonance, which occurs at $\theta=\pi/2$, the polarizability is purely imaginary and has the value

 \begin{equation}
     \alpha_m = \frac{-j}{r(\omega_0)}.
 \end{equation}

 \noindent At resonance we can determine the total power radiated using 

 \begin{equation}
    P_{\textrm{abs}}=\frac{\mu \omega}{2}\textrm{Im}\{ \alpha_m \} |H_0^+|^2,
    \label{eq:power_absorbed}
\end{equation}

\noindent where $H_0^+$ is the incident waveguide mode. An expression for $H_0^2$ can be found as \cite{smith2024circuit}

\begin{equation}
    |H_0^+|^2=\frac{2}{a b Z_0^2},
\end{equation}

\noindent so that we have

\begin{equation}
    P_{\textrm{rad}}(\omega_0) = \frac{\mu \omega}{a b Z_0^2}\frac{1}{r(\omega_0)}.
\end{equation}

\noindent This final expression shows that the total radiated power is entirely independent of the characteristics of the aperture, just as is the polarizability. The constancy of the radiated power for any resonant aperture enables us to determine an approximate expression for the transformer ratio of the equivalent circuit in Fig. \ref{fig:single_cELC}(a), as is presented in the appendix.

\subsection{Inclusion of Resistive and Dielectric Losses}

In addition to radiative losses there are two other potential loss mechanisms for the cELC. Since the metal forming the waveguide structure will have finite conductivity, there will be a resistance to be included in the circuit, in addition to the inductance and capacitance due to resistive losses. Furthermore, if the cELC is constructed using circuit board (as is common) or other substrate materials, dielectric losses will be present. For a complete characterization of the element, both sources of loss should be included. In this section, we neglect propagation loss within the waveguide structure and assume the resistive and dielectric losses are in the immediate vicinity of the metamaterial element. That is, we assume losses only in the regions where either the electric fields of the resonant element are enhanced, or where the local currents are strongly perturbed.

Because the non-radiative loss can be expected to be important mainly near the resonance frequency, we assume $\omega^2\ll\omega_1^2$ so that the polarizability in (\ref{eq:celc_polarizability}) reduces to

\begin{equation}
    \alpha_m = \frac{\alpha_{m0} \omega_0^2}{\omega_0^2-\omega^2+j \alpha_{m0} \omega_0^2 \Big( \frac{\beta}{ab}+\frac{k^3}{3\pi}\Big)}
    \label{eq:polarizability_near_resonance}
\end{equation}

We first account for the metal losses by assuming a resistance that is in series with the inductance. We can then replace the internal and external inductances by,

\begin{equation}
    L'_i \rightarrow L'_i + \frac{R'_i}{j \omega},
\end{equation}

\noindent and

\begin{equation}
    L_e \rightarrow L_e + \frac{R_e}{j \omega},
\end{equation}

\noindent where we assume

\begin{equation}
    \frac{R'_i}{j\omega L'_i}\ll 1,
\end{equation}

\noindent and

\begin{equation}
    \frac{R_e}{j\omega L_e}\ll 1.
\end{equation}

\noindent Since the losses are added perturbatively, we can ignore their effects in the numerator of the impedance, i.e., (\ref{eq:celc_impedance}), including the loss terms only in the denominator. With the loss terms included, the resonance frequency can be written

\begin{equation}
    \omega_0^{\prime 2} = \omega_0^2 \frac{1}{1+\frac{\Gamma_m}{j \omega}},
\end{equation}

\noindent where

\begin{equation}
    \Gamma_m = \frac{R'_i+R_e}{L'_i+L_e}.
\end{equation}

\noindent With metal losses included, the final polarizability has the form

\begin{equation}
    \alpha_m = \frac{\alpha_{m0} \omega_0^2}{\omega_0^2-\omega^2+j \Big[ \omega \Gamma_m+ \alpha_{m0} \omega_0^2 \Big( \frac{\beta}{ab}+\frac{k^3}{3\pi}\Big) \Big]}   
\end{equation}

Dielectric loss can be included in a similar manner, although the situation is more complicated as only a portion of the electric flux will flow through the dielectric regions that are within the waveguide or near the cELC. Assuming there is a dielectric material very near the high-capacitive regions of the cELC and that the bulk of the electric flux flows through this region, then dielectric loss can potentially be approximated by adding a conductance in parallel with the capacitance, such that

\begin{equation}
    C'_i \rightarrow C'_i+\frac{G}{j \omega}.
    \label{eq:capacitance_with_loss}
\end{equation}

\noindent Following the same steps as for the case of metal losses, the final polarizability is

\begin{equation}
    \alpha_m = \frac{\alpha_{m0} \omega_0^2}{\omega_0^2-\omega^2+j \Big[ \omega \Gamma_d+ \alpha_{m0} \omega_0^2 \Big( \frac{\beta}{ab}+\frac{k^3}{3\pi}\Big) \Big]},   
\end{equation}

\noindent where

\begin{equation}
    \Gamma_d = \frac{G}{C'_i}.
\end{equation}

\noindent This expression is likely unreliable, however. A more appropriate approach would be to divide the capacitance into a part due to the free space region and a second part due to the dielectric region, or $C'_i = C'_{i,f}+C'_{i,d}$, replacing (\ref{eq:capacitance_with_loss}) with $C'_{i,d}$. A more accurate circuit could also likely be derived from a small aperture coupling analysis that includes different media on either side of the iris \cite{liang1983generalized, mautz1983admittance}.

Using (\ref{eq:power_absorbed}), we obtain

\begin{equation}
    \bar{P}_{\textrm{abs}}=\frac{2\beta \alpha_{m0}}{a b} 
    \frac{\omega_0^2 \omega \Gamma}
    {(\omega_0^2-\omega^2)^2+\Big[ \omega \Gamma + \alpha_{m0} \omega_0^2 \Big( \frac{\beta}{a b}+\frac{k^3}{3 \pi}
    \Big) \Big]^2}.
\end{equation}

\noindent Here, $\Gamma$ refers to either the metal or the dielectric damping factor. If both losses are present and are relatively small, then $\Gamma \approx \Gamma_m + \Gamma_e$.

The power absorbed due to losses is in addition to the power radiated, which can be found from

\begin{equation}
    \bar{P}_{\textrm{rad}}=\frac{2\beta }{a b} 
    \frac{(\alpha_{m0} \omega_0^2)^2 \frac{k^3}{3 \pi}}
    {(\omega_0^2-\omega^2)^2+\Big[ \omega \Gamma + \alpha_{m0} \omega_0^2 \Big( \frac{\beta}{a b}+\frac{k^3}{3 \pi}
    \Big) \Big]^2}.
\end{equation}

\noindent The ratio of the radiated power to dissipated power is then

\begin{equation}
    \frac{\bar{P}_{\textrm{rad}}}{\bar{P}_{\textrm{abs}}}=
    \frac{\alpha_{m0}\omega_0^2 k^3}{3 \pi \omega \Gamma}
\end{equation}

Note that in the absence of absorptive loss, the imaginary part of the polarizability at resonance is determined entirely by the radiation damping terms, or

\begin{equation}
    |\textrm{Im}(\alpha_m(\omega_0))|=\frac{1}{\frac{\beta}{a b}+\frac{k^3}{3 \pi}}=\alpha_i.
\end{equation}

\noindent In the case where absorptive loss is present, then

\begin{equation}
    |\textrm{Im}(\alpha_m(\omega_0))|=\frac{\alpha_{m0} \omega_0^2}{\omega_0 \Gamma + \alpha_{m0} \omega_0^2 \Big( \frac{\beta}{a b}+\frac{k^3}{3 \pi}\Big)}.    
\end{equation}

\noindent or

\begin{equation}
    |\alpha^{\prime \prime}|=\frac{\alpha_{m0}}{\frac{ \Gamma}{\omega_0} + \frac{\alpha_{m0}}{\alpha_i}}=
    \frac{\omega_0}{\frac{\Gamma}{\alpha_{m0}} + \frac{\omega_0}{\alpha_i}}.    
\end{equation}

\noindent Since the imaginary part of the polarizability can be obtained from measurement or simulation, the above equation allows the ratio $\Gamma/\alpha_{m0}$ to be obtained as

\begin{equation}
    \frac{\Gamma}{\alpha_{m0}}=\omega_0 \Big( \frac{1}{|\alpha^{\prime \prime}|}-\frac{1}{\alpha_i}\Big)
\end{equation}

\noindent We can obtain a second measurement to separately obtain $\Gamma$ and $\alpha_{m0}$ by noting the point where $\textrm{Re}(\alpha_m(\omega_r))=-\textrm{Im}(\alpha_m(\omega_r))$, which yields

\begin{equation}
    \alpha_{m0}=\frac{\omega_r^2-\omega_0^2}
    {\omega_r \frac{\Gamma}{\alpha_{m0}}+\omega_0^2 \Big( \frac{\beta(\omega_r)}{a b} + \frac{k(\omega_r)^3}{3 \pi}\Big)}.
\end{equation}

\noindent Thus, from two measurements of the extracted polarizability, the parameters $\alpha_{m0}$ and $\Gamma$ can be found. This procedure allows separation of the radiative and absorptive losses without separate measurement or consideration of the radiated power.

\subsection{cELC Polarizability and Circuit Retrieval}

\begin{figure}[!b]
\centering
\includegraphics[width=2.5in]{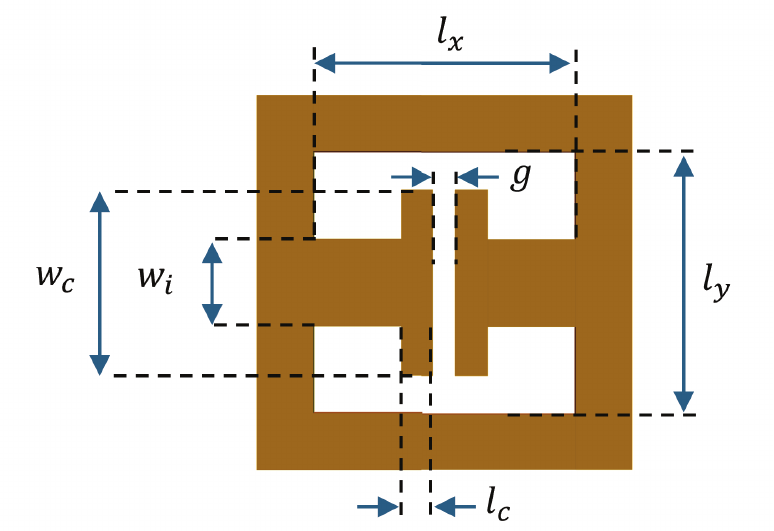}
\caption{A typical cELC design and relevant geometrical parameters.}
\label{fig:cELC_dimensions}
\end{figure}

To demonstrate the applicability of the cELC circuit model, we demonstrate the steps of polarizability extraction and equivalent circuit retrieval for a sample cELC element, such as that shown in Fig. \ref{fig:cELC_dimensions}. The cELC has been designed so as to resonate at a frequency within the X-band (8-12 GHz), where the rectangular waveguide is single mode. The height and width of the waveguide are $b=0.5$ cm and $a=2.29$ cm, respectively, with a metal thickness of $t=0.2$ mm. The metal here is modeled as a perfect electrical conductor. The waveguide is assumed to be hollow, so that the interior dielectric constant is unity. If the interior of the waveguide were filled with a dielectric material such that the regions inside and outside the waveguide were dissimilar, the circuit model would need to be appropriately modified based on a modified aperture model \cite{mautz1983admittance, liang1983generalized}. We defer such an analysis for future studies. For this first example, the various geometrical parameters of the cELC were chosen arbitrarily and are indicated in Table \ref{table:celc_parameters_1}, v1. 

\renewcommand{\arraystretch}{1.2}
\begin{table}
\caption{Geometrical parameters for the cELC elements}
\small
\centering
\begin{tabular}{ |p{1.5cm}|p{1.5cm}|p{1.5cm}|p{1.5cm}|  }
\hline
\multicolumn{4}{|c|}{Parameters (units in mm)}\\
\hline
Parameter & v1 & v2 & v3 \\
\hline
$l_x$ & 4.8 & 4.8 & 4.8 \\
$l_y$ & 4.8 & 3.2 & 5.2 \\
$w_i$ & 1.6 & 0.2 & 0.2 \\
$w_c$ & 3.4 & 1.8 & 2.8 \\
$l_c$ & 1.2 & 0.51 & 0.3 \\
$g$ & 0.4 & 0.18 & 0.6 \\
\hline
\end{tabular}
\vspace*{2mm}
\label{table:celc_parameters_1}
\end{table}

\begin{figure}[!b]
\centering
\includegraphics[width=3.5in]{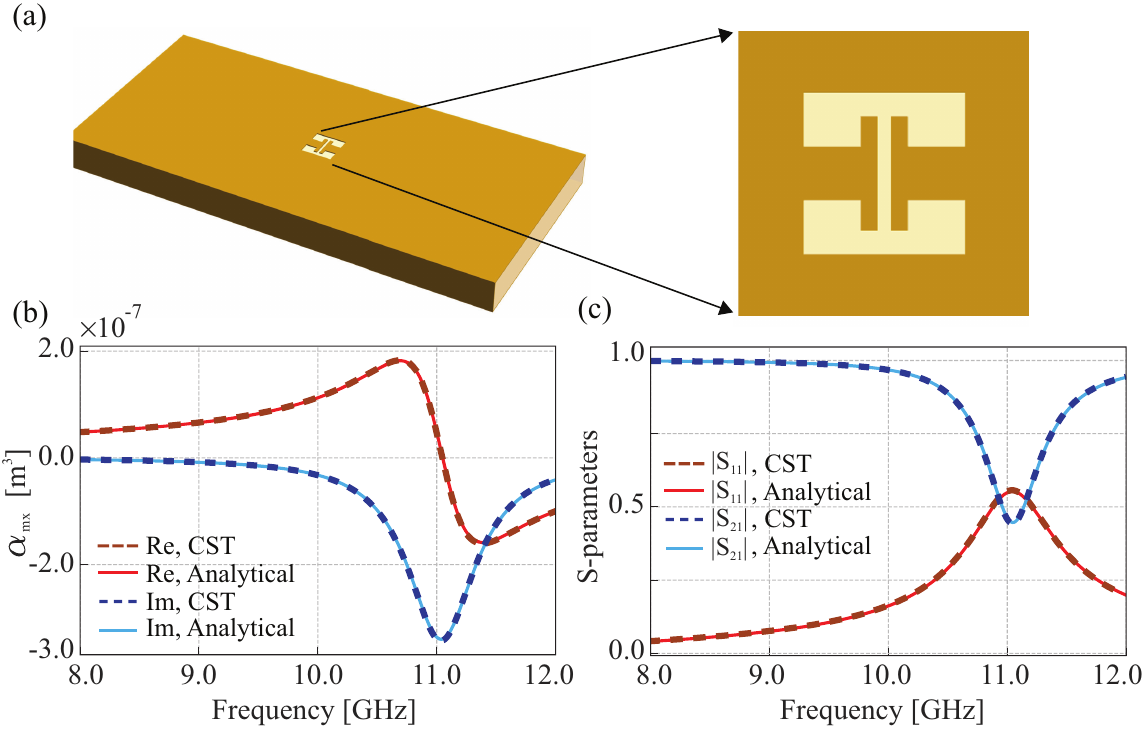}
\caption{A cELC element designed to exhibit a resonance in the X-band range. (Top) Perspective and cELC detail. (Left) Numerically computed (dashed) and analytical (solid) extracted polarizability. (Right) S-parameters, numerically computed (dashed) and analytically calculated (solid).}
\label{fig:cELC_Example_1}
\end{figure}

The effective electric and magnetic polarizabilities of the metamaterial element can be extracted from the computed waveguide scattering (S-) parameters by use of the following equations \cite{pulido2017polarizability,smith2024circuit}:

\begin{equation}
\label{eq:polarizabilities_vs_s_parameters}
\begin{aligned}
    \alpha_{ey} &=\frac{j a b \beta}{2 k^2}\left(S_{21}+S_{11}-1\right), \\
    \alpha_{mx} &=\frac{j a b}{2 \beta}\left(S_{21}-S_{11}-1\right).
\end{aligned}
\end{equation}

\noindent For the results presented here, the full-wave solver Microwave Studio (CST) was used to compute the S-parameters for simulated cELC structures, which were subsequently exported as text files for further postprocessing using a custom Python script. An approximation assumed here is that the response of the cELC element produces only a magnetic polarizability, with the electric polarizability being neglected. Comparing the magnitudes of the extracted electric and magnetic polarizabilities for this and subsequent examples, it is confirmed that this approximation holds and thus the electric polarizability is neglected in the ensuing analysis.

The effective circuit parameters can be determined by performing a multi-variable fit to the polarizability extracted from the computed S-parameters. From the analysis above, we find that combinations of the relevant circuit parameters contribute to three parameters within the polarizability: $\alpha_{m0}$, $\omega_0^2$, and $\omega_1^2$, as can be observed from (\ref{eq:celc_polarizability}). The polarizability expression, however, also involves the dispersive radiation damping terms that would appear to complicate the fitting unless included explicitly.

To avoid the complications associated with fitting to the polarizability, we consider its inverse, noting the following:

\begin{equation}
    \textrm{Re} \Big\{ \frac{1}{\alpha_m} \Big\} = \frac{1}{\alpha_{m0}}\frac{1-\omega^2/\omega_0^2}{1-\omega^2/\omega_1^2}
    \label{eq:real_inverse_polarizability}
\end{equation}

\noindent and

\begin{equation}
    \textrm{Im} \Big\{ \frac{1}{\alpha_m} \Big\} = \frac{\beta}{ab}+\frac{k^3}{3\pi}
    \label{eq:imag_inverse_polarizability}
\end{equation}

Conveniently, when the expression $1/\alpha_m$ is used, the radiative damping terms can be separated from the reactive components. This division is possible because the radiation damping terms equate to a parallel resistance in the circuit model, such that the equivalent admittance is a sum of the susceptive and conductive contributions.

From (\ref{eq:real_inverse_polarizability}), it can be seen that the three fitting parameters can be found from the real part of the inverse of the polarizability. Moreover, the resonance frequency enters the equation as a zero, while the second frequency of interest enters as a pole---both easily resolvable via a curve fit. Finally, the inverse of the static polarizability enters as a simple constant that corresponds to the strength of the resonance.

To determine the equivalent circuit parameters of the cELC in Fig. \ref{fig:cELC_Example_1}, the polarizability was first extracted, followed by performing a fit to the real part of $1/\alpha_m$. The static polarizability, $\omega_0^2$, and $\omega_1^2$ were then found and used to arrive at values for $L'_i$, $L_e$, and $C'_i$. The resulting parameters are shown in Table \ref{table:celc_fit_pars}. (The values of the parasitic inductance, $L_p$, are associated with a lumped capacitor added to the circuit and are not inherent to the cELC.)

\renewcommand{\arraystretch}{1.2}
\begin{table}
\caption{Extracted Parameters for the cELC Elements}
\small
\centering
\begin{tabular}{ |p{0.8 cm}|p{1.cm}|p{1.cm}|p{1.cm}|p{1.5cm}|  }
\hline
\multicolumn{5}{|c|}{Fit Parameters}\\
\hline
Pars & v1 & v2 & v3 & Units \\
\hline
$L_{e}$ & 542.2 & 223.0 & 667.3 & pH \\
$L'_{i}$ & 79.1 & 131 & 382.3 & pH \\
$C'_{i}$ & 0.33 & 0.58 & 0.20 & pF \\
$n^2$ & 5.15 & 6.8 & 4.03 & - \\
\hline
$\alpha_{m0}$ & 2.47 & 1.02 & 3.04 & $\times 10^{-8} \textrm{m}^3$ \\
$f_0$ & 11.04 & 11.11 & 10.93 & GHz \\
$f_1$ & 30.94 & 18.26 & 18.11 & GHz \\
\hline
$L_{p}^{*}$ & $127.3^{*}$ & $40^{*}$ & $385^{*}$ & pH \\
\hline
\end{tabular}
\vspace*{2mm}
\label{table:celc_fit_pars}
\end{table}

The simulated and fitted polarizabilities and S-parameters are shown in Fig. \ref{fig:cELC_Example_1}. As can be seen from the figure, the agreement between the computed and theoretically determined curves is excellent. The agreement is particularly striking since the width and magnitude of the resonance---determined by the radiative damping factors---are not considered in the fitting. The radiation damping terms are entirely analytical, yet lead to near exact agreement with the simulated curves.

\begin{figure}[!b]
\centering
\includegraphics[width=3.35in]{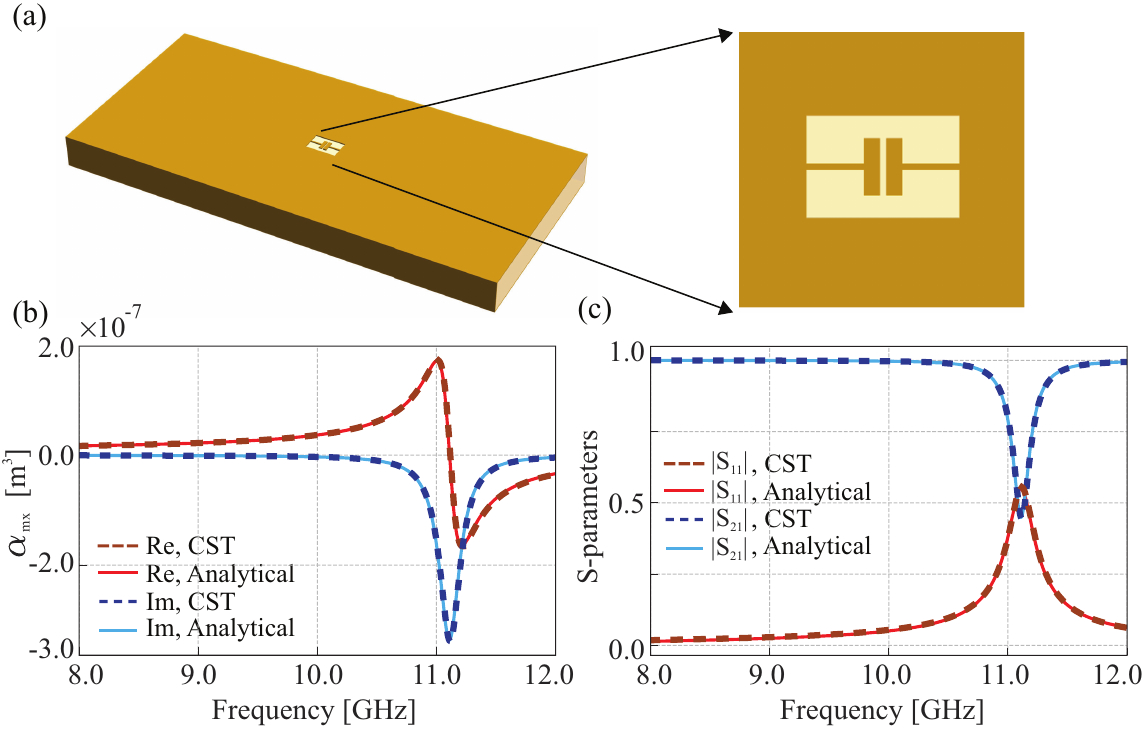}
\caption{cELC element v2 designed to exhibit a resonance in the X-band range. (Top) Perspective and cELC detail. (Left) Extracted real and imaginary polarizability (dashed curves) and the analytically determined polarizability (solid curves). (Right) Simulated $S_{11}$ and $S_{21}$ (dashed curves) computed and the analytically determined S-parameters (solid curves).}
\label{fig:cELC_Example_2}
\end{figure}

\subsection{Modifying the cELC Bandwidth}

The effective Q-factor of the polarizability resonance is determined largely by the static polarizability, $\alpha_{m0}$, which in turn depends on the inductance of the element. Thus, if the resonance frequency is held constant for a given design, the Q-factor can be decreased by minimizing the inductance of the element and compensating with additional capacitance, or increased by doing the reverse. To illustrate this effect, we design two cELC elements that should display different Q-factors. As a first example, we decrease the width of the element, $l_y$, which reduces the external inductance. We then decrease the gap between the two center arms, $g$, to increase the capacitance, aiming for a resonance frequency near 11 GHz. The resulting element is shown in Fig. \ref{fig:cELC_Example_2}, with its geometrical parameters summarized in Table \ref{table:celc_parameters_1}, v2.

As can be seen in Fig. \ref{fig:cELC_Example_2}, the Q-factor is relatively sharper for v2 than for v1. The extracted circuit and polarizability parameters reveal the expected behavior, with the external inductance decreasing to 223 pH (from 542 pH for v1) and the internal capacitance increasing to 0.58 pF (from 0.33 pF for v1). As expected, the extracted static polarizability is reduced to $\alpha_{m0}=1.02 \times 10^{-8} \textrm{ m}^3$ (from $\alpha_{m0}=2.47 \times 10^{-8} \textrm{ m}^3$ for v1). As before, the curves corresponding to the simulated S-parameters and extracted polarizabilities (dashed curves) are compared with the analytical expressions (solid curves); the agreement is so close that they essentially lie on top of each other.

\begin{figure}[!b]
\centering
\includegraphics[width=3.35in]{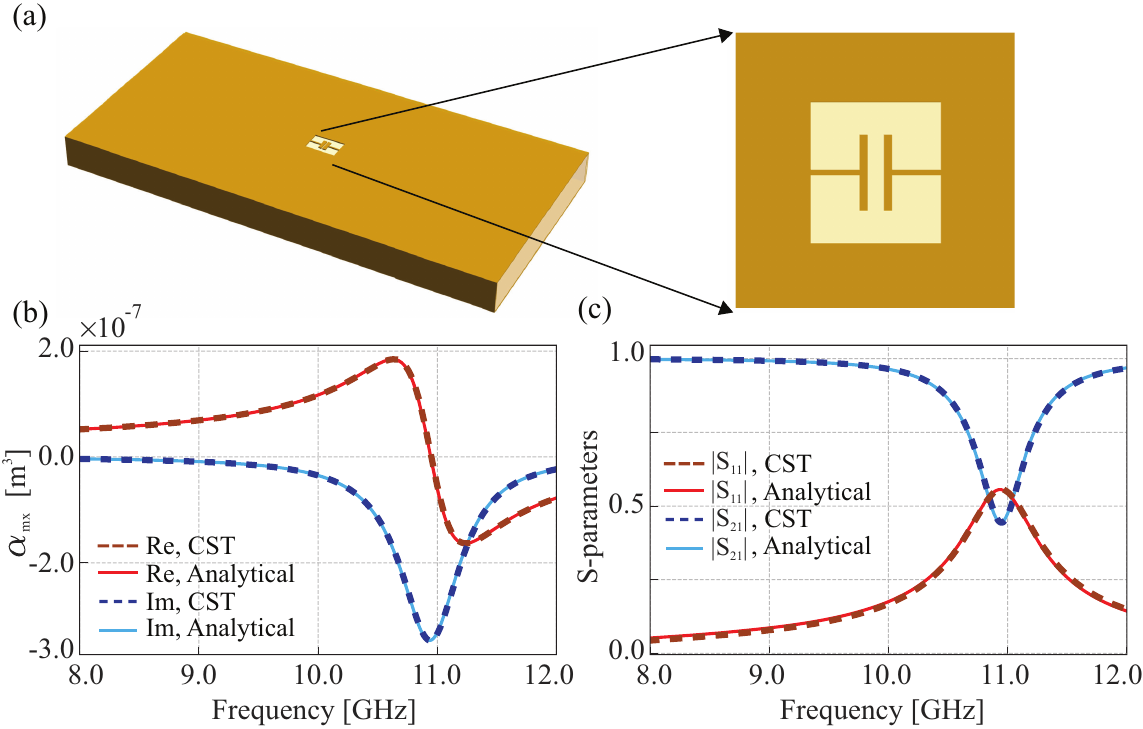}
\caption{cELC element v3 designed to exhibit a resonance in the X-band range. (Top) Perspective and cELC detail. (Left) Extracted real and imaginary polarizability (dashed curves) and the fitted polarizability (solid curves). (Right) Simulated $S_{11}$  and $S_{21}$ (dashed curves) computed and the fitted S-parameters (solid curves).}
\label{fig:cELC_Example_3}
\end{figure}

In a similar manner, the geometry of the element can be adjusted so as to decrease the Q-factor by reducing the capacitive contribution and increasing the inductive contributions. An example is shown in Fig. \ref{fig:cELC_Example_3}, where the width of the element has been increased, the thicknesses of the central conductors have been decreased, and the gap between conductors increased. The geometrical parameters are shown in Table \ref{table:celc_parameters_1}, v3. The resulting polarizability and S-parameter curves, shown in Fig. \ref{fig:cELC_Example_3}, reveal the substantially lower Q-factor. As before, both the simulated and analytical curves are plotted for the polarizability and S-parameters, with excellent agreement found between the two.

As predicted by the analytical results, the peak polarizability (at resonance) is identical for all elements (i.e., v1, v2 and v3); likewise, the peak of $S_{11}$ and minimum of $S_{21}$ are also identical for the three examples. Regardless of the cELC properties, the maximum polarizability and scattering are the same and set entirely by the radiation damping terms. Only the width of the resonance is adjusted by changing the strength of the polarizability. This result shows that the coupling of slot-like elements and apertures to the waveguide feed is immutable, tied to the Green's functions for the waveguide and free space. Note that this coupling can be adjusted by moving the element away from the center axis of the waveguide for rectangular waveguides, where the transverse magnetic fields are significantly reduced near the waveguide walls. Similarly, waveguides filled with different dielectric materials, or different waveguide types will generally modify the coupling.

\subsection{The Disconnected cELC}

\begin{figure}[!b]
\centering
\includegraphics[width=2.5in]{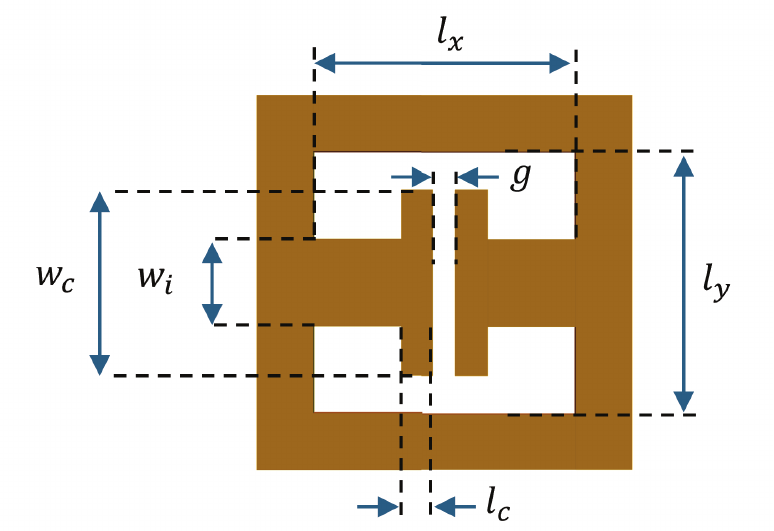}
\caption{The disconnected cELC design and relevant geometrical parameters.}
\label{fig:disconnected_cELC_dimensions}
\end{figure}

The cELC design presented in the previous sections is characterized by a single metal surface. While this design is useful to develop the basic theory of the cELC resonator, it is more difficult to implement in practice when external elements are added that require bias. For dynamically reconfigurable metasurfaces, a far more common design is that of the disconnected cELC, which consists of two disconnected conducting surfaces \cite{sleasman2016waveguide,boyarsky2021electronically,yoo2023experimental}. Since the two sections of the cELC are isolated, a voltage bias can easily be applied to the element that will provide DC control for elements such as diodes or varactors integrated into the cELC gaps. A depiction of the disconnected cELC geometry is shown in Fig. \ref{fig:disconnected_cELC_dimensions}, along with relevant geometrical parameters.

If the disconnected cELC is to be used in a symmetric manner, there is an inherent complication that arises; there are now two capacitive gaps that will contribute to the overall capacitance. The total capacitance, then, will be half the capacitance associated with either gap alone, which will move the resonance to higher frequencies. For this reason, the disconnected cELC is often used with the added capacitors (or varactors) providing the main source of capacitance, rather than relying on the geometrical capacitance.

An additional complication exists for the disconnected cELC in that there are effectively two magnetic currents that radiate as two magnetic dipoles. Assuming these slots radiate equally, then there will be a slight modification to the free space and waveguide radiation patterns leading to slightly different radiation damping factors. If the ratio of these factors is modified, then there will be a modification required for the effective polarizability.

\begin{figure}[!b]
\centering
\includegraphics[width=3.15in]{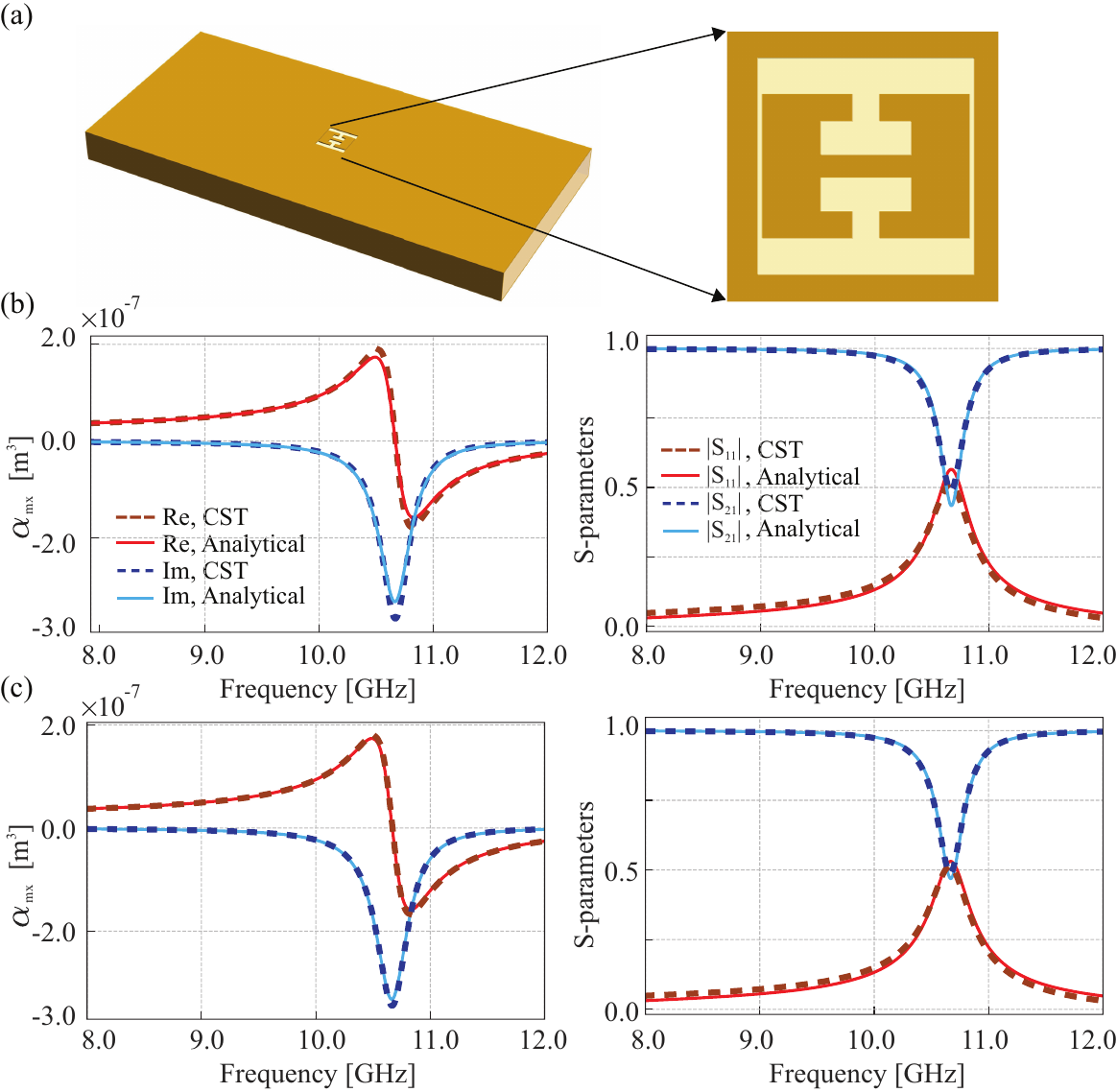}
\caption{Disconnected cELC element designed to exhibit a resonance in the X-band range. (a) Perspective and cELC detail. (b), (c) Extracted real and imaginary polarizability (dashed curves) and the fitted polarizability (solid curves), and simulated $S_{11}$ and $S_{21}$ (dashed curves) computed and the fitted S-parameters (solid curves).}
\label{fig:disc_celc_composite}
\end{figure}

As an example, we simulate a disconnected cELC of the form shown in Fig. \ref{fig:disc_celc_composite}. The parameters for this element are chosen as $l_x=l_y=4.8$ mm, $s_x=0.1$ mm, $s_y=$0.8 mm, $t_o=0.6$ mm, $t_i=0.5$ mm, $g_o=0.6$ mm, and $g_i=2.0$ mm. The resulting polarizability and S-parameter curves are shown in Fig. \ref{fig:disc_celc_composite} (dashed curves), along with their curve fits (solid curves). The agreement over most of the curves is quite good, except near the resonance. Recall that at the resonance all of the reactive factors cancel out, leaving the peak polarizability set by the radiation damping terms. The disagreement between the extracted and fitted polarizabilities at and near resonance suggests that the ratio of the radiative damping terms has been modified by the presence of the two radiating dipoles. Rather than attempting to compute the correction, we instead empirically multiply the free-space radiation damping term by a factor of $1.15$ to account for this difference. The corrected extracted and fitted curves are shown in Fig. \ref{fig:disc_celc_composite}, where it can be seen the agreement is now restored. The circuit parameters determined for this element are $L_e=522$ pH, $L'_i=670$ pH, and a total capacitance of $C'_i=0.186$ pF. Thus, each gap would be expected to have a capacitance of $0.372$ pF.


One other modification that potentially arises for the disconnected cELC is that the spatial speparation between the radiating slots implies the excitation fields may be different at the two slots. This situation is possible if the interaction between the slots is strong enough, inducing an electromagnetic asymmetry to the otherwise geometrically symmetric element. This particular interaction may be evident from the residual disagreement between the $S_{11}$ curves in Fig. \ref{fig:disc_celc_composite}. While the impact is relatively small on the scattering parameters and virtually nonexistent in the corrected polarizability curves, nevertheless there may be an impact on the radiated field distribution that would need to be addressed. Taking into account the dipole interactions between the slots through a coupled-dipole formalism may address all of these complications, but is beyond the scope of this current work and will be left for future studies.

The disconnected cELC was purposefully designed to be resonant within the band of interest, which required extremely narrow gaps between the two conductors. For dynamically tunable cELCs, typically, the intrinsic capacitance of the disconnected cELC would be reduced by making the gaps larger, such that the element would be resonant at much higher frequencies. The addition of one or more lumped capacitive elements would then define the actual resonance frequency. To apply the theory developed here, a modified fitting procedure would need to be followed that would recover the resonance and zero frequencies.

\section{Tuning cELC with Lumped Elements}
\label{sec:celc_tuning}

\begin{figure}[!b]
\centering
\includegraphics[width=3.0in]{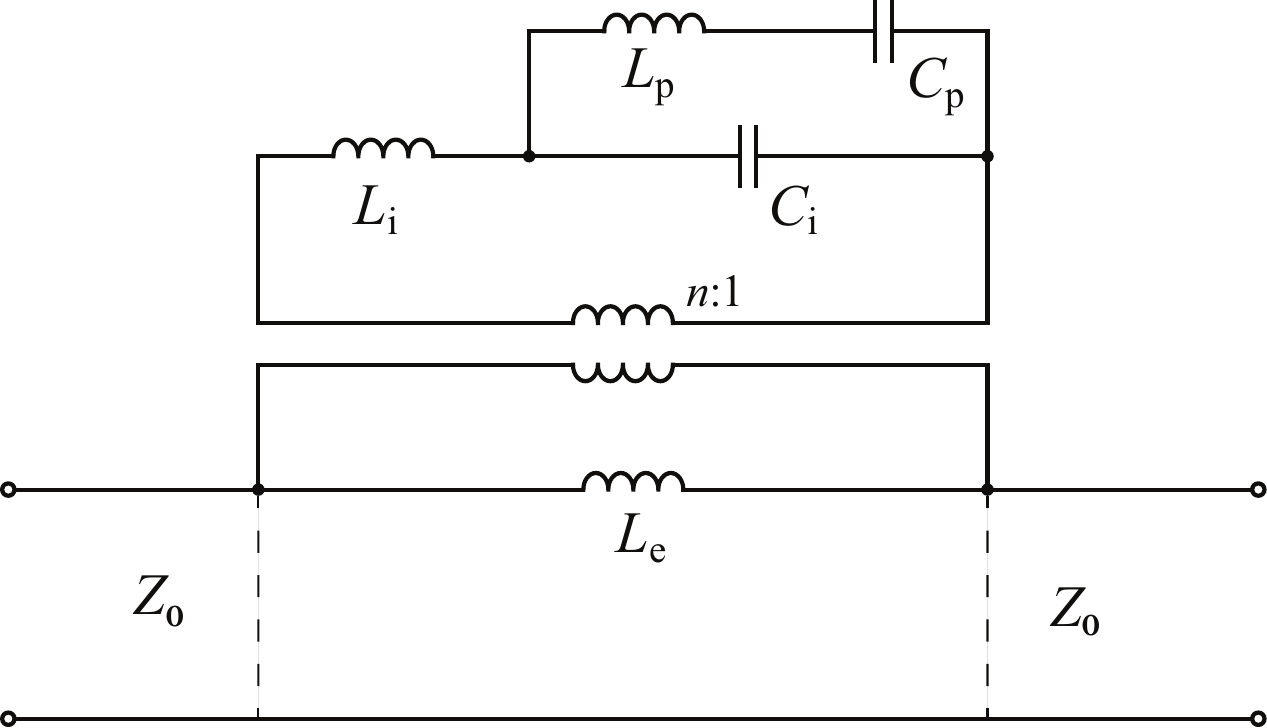}
\caption{Circuit diagram of a cELC with capacitor integrated into the slot region. In this model, a package inductance is included in the integrated capacitor for generality.}
\label{fig:cELC_with_varactor}
\end{figure}

The cELC was one of the first elements to be incorporated into a dynamically reconfigurable waveguide-fed metasurface, with diodes or varactors integrated into the capacitive gaps and connected to an external voltage source \cite{sleasman2016waveguide,boyarsky2021electronically,yoo2023experimental}. The addition of an externally controlled, variable capacitive element to the gap region of an element enables the resonance frequency of the element to be tuned via application of an external bias voltage. Having developed a reasonable circuit model for the cELC, we next extend this model to include lumped elements. The lumped elements may be externally tunable, such as diodes or varactors, but we simplify the discussion here by restricting our attention to a passive capacitor, avoiding the discussion of additional bias circuitry that would entail more design considerations for the metamaterial element. The capacitor-loaded cELC proposed circuit model is shown in Fig. \ref{fig:cELC_with_varactor}, where the added capacitor $C'_p$ is placed across the slot capacitance $C'_i$. For the model to be general, an additional parasitic inductance is added in series with the external capacitor, $L'_p$, since any inserted device will necessarily possess some amount of parasitic inductance due to its packaging and finite length of leads. Note that the equivalent circuit of Fig. \ref{fig:cELC_with_varactor} includes the transformer coupling; however, in the following we assume the impedances have all been transformed so that this complication can be ignored. 

The impedance of the model of Fig. \ref{fig:cELC_with_varactor} can be computed as

\begin{equation}
    Z=j \omega L_e \frac{\omega_0^2}{\omega_1^2}\Big[\frac
    {\omega_1^2 \omega_p^2 -(\omega_1^2+\omega_p^2)\omega^2+f\omega^4}
    {\omega_0^2 \omega_p^2 -(\omega_0^2+\omega_p^2)\omega^2+f\omega^4}
    \Big],
    \label{eq:z_of_celc_with_cap}
\end{equation}

\noindent where

\begin{equation}
    \omega_0^2 = \frac{1}{(L'_i+L_e)(C'_i+C'_p)},
\end{equation}

\begin{equation}
    \omega_1^2 = \frac{1}{L'_i(C'_i+C'_p)},
\end{equation}

\begin{equation}
    \omega_p^2 = \frac{1}{L'_p C'_p},
\end{equation}

\noindent and

\begin{equation}
    f=\frac{C'_i}{C'_i+C'_p}.
\end{equation}

\noindent The impedance has been written in such a manner that, in the limit that the parasitic inductance is very small, $\omega_p \rightarrow \infty$, so that the resonance frequency can be immediately seen as $\omega_0$. In this limit, the only change to the previous cELC model is that the external capacitance simply adds to the slot capacitance. The model is then trivial to implement.

\subsection{Estimating the Parasitic Inductance}
The impedance, in its general form, possesses two poles and two zeros for $\omega^2$. While the previous method of fitting could, in principle, recover all five unknown circuit elements (i.e., $L_e$ and the two poles and two zeros), such a fitting is complicated since at least one of the poles and one of the zeros occur at much higher frequencies than the frequency band of interest. In addition, for waveguide-fed elements, these frequencies will lie beyond the cutoff frequency of the next waveguide mode, so that even broad frequency-span simulations may be complicated due to higher-order mode excitation. To avoid these complications, we instead perform a second simulation in which the added capacitance $C_p$ is made very large, enabling us to determine $L_p$. For the examples presented here (v2 and v3), we use a value of $C_p=50$ pF. In the limit $C_p \rightarrow \infty$ it can be seen all of the frequency parameters $\omega_0^2$, $\omega_1^2$ and $\omega_p^2$ and $f$ tend to zero, so that the impedance reduces to

\begin{equation}
    Z=j \omega L_e \Bigg[ \frac{(1+\frac{\omega_p^2}{\omega_1^2})-\frac{f}{\omega_1^2}\omega^2}{(1+\frac{\omega_p^2}{\omega_0^2})-\frac{f}{\omega_0^2}\omega^2} \Bigg],
    \label{eq:impedance_celc_with_cap}
\end{equation}

\noindent which can be written

\begin{equation}
    Z=j \omega L_e \frac{L'_i+L_p}{L'_i+L_e+L'_p} \frac{1-\omega^2/\omega_1^{'2}}{1-\omega^2/\omega_0^{'2}},
\end{equation}

\noindent where

\begin{equation}
    \omega_0^{'2}=\frac{1}{[L'_p \parallel (L'_i+L_e)]C'_i}
\end{equation}

\begin{equation}
    \omega_1^{'2}=\frac{1}{(L'_p \parallel L'_i)C_i}.
\end{equation}

For the shorted cELC, then, the polarizability thus takes the form

\begin{equation}
    \alpha_m = \frac{\alpha'_{m0}(1-\omega^2/\omega_1^{'2})}{(1-\omega^2/\omega_0^{'2})+j\alpha'_{m0}(1-\omega^2/\omega_1^{'2})\Big( \frac{\beta}{ab}+\frac{k^3}{3\pi}\Big)},
    \label{eq:shorted_celc_polarizability}
\end{equation}

\noindent with

\begin{equation}
    \alpha'_{m0} = \frac{ab}{2\mu} \frac{L_e(L'_i+L'p)}{L'_i+L_e+L'_p}.    
\end{equation}

\noindent Our strategy for estimating the parasitic inductance for the shorted cELC is to first extract the polarizability, then perform a fit to (\ref{eq:shorted_celc_polarizability}) to determine the coefficient $\alpha'_{m0}$. Since the resonance frequency $\omega'_0$ is moved to much lower frequencies---out of the band of interest---the slot is nonresonant and the damping terms can be neglected. Thus, we can set the damping terms equal to zero and attempt a fit to 
$\alpha_{m}$ to ultimately extract $L'_p$.

This fitting procedure, followed for the three cELC designs, produced the values of $L'_p$ shown in Table \ref{table:celc_fit_pars}. The obtained parasitic inductances are consistent with intuition; for example, the cELC with the smallest interior gap has the smallest value of $L'_p$.

\subsection{cELC Tuning}

With the circuit parameters determined, the impact of adding a lumped capacitor to the capacitive gap region within the cELC can be immediately determined using (\ref{eq:z_of_celc_with_cap}). Were the transformer turns ratio $n^2$ equal to unity and the added capacitance relatively small, then it would be anticipated that the increase in capacitance would be simply the sum of the geometrical capacitance and that from the added lumped element. However, near resonance $n^2>1$, so that the physical value of the added capacitance must be multiplied by $n^2$ to find the effective value $C'_p$. As described in the appendix, an expression for $n^2$ can be determined and is presented in Table \ref{table:celc_fit_pars}. These values are approximate, but provide some indication of the expected impact of the external capacitor on the resonance.  

\begin{figure}[!b]
\centering
\includegraphics[width=3.5in]{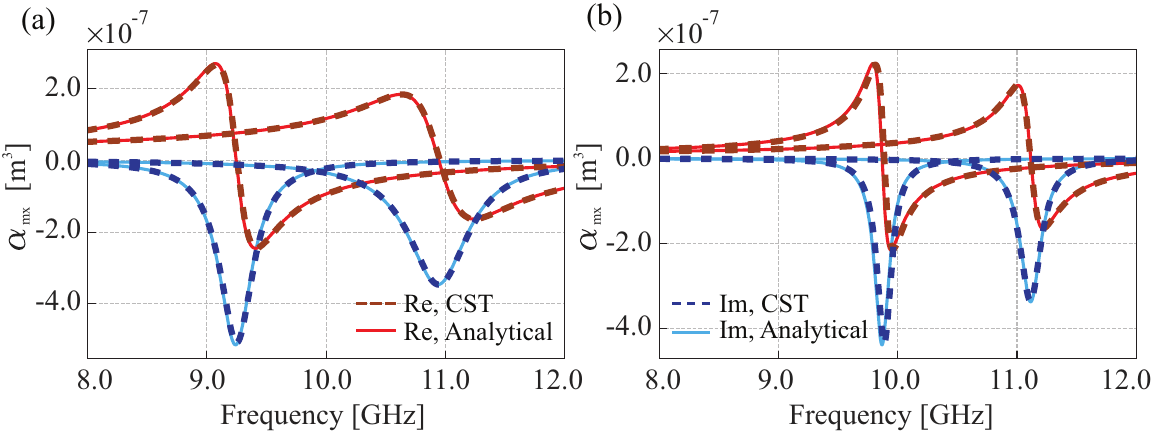}
\caption{(a) Polarizability curves for the v3 cELC with and without an added 0.02 pF capacitor. The bare cELC is shown as the polarizability curve with resonance near 11 GHz, while the tuned cELC has been shifted to about 9.2 GHz. The simulated (dashed) and fitted (solid) curves are shown. (b) Same curves for the v2 cELC.}
\label{fig:cELC_tuning_examples}
\end{figure}

We illustrate tuning the cELC via full-wave simulations in CST on the v2 and v3 elements by adding a lumped capacitance of 0.02 pF. The results are shown in Fig. \ref{fig:cELC_tuning_examples}. In this simulation, the cELC was initially shorted with a 50 pF capacitor to determine $L'_p$, as described in the previous section. The rather modest value of $C_p=0.02$ pF is seen to shift the resonance of the elements considerably, across the X-band frequency span. 

\begin{figure}[!b]
\centering
\includegraphics[width=3.5in]{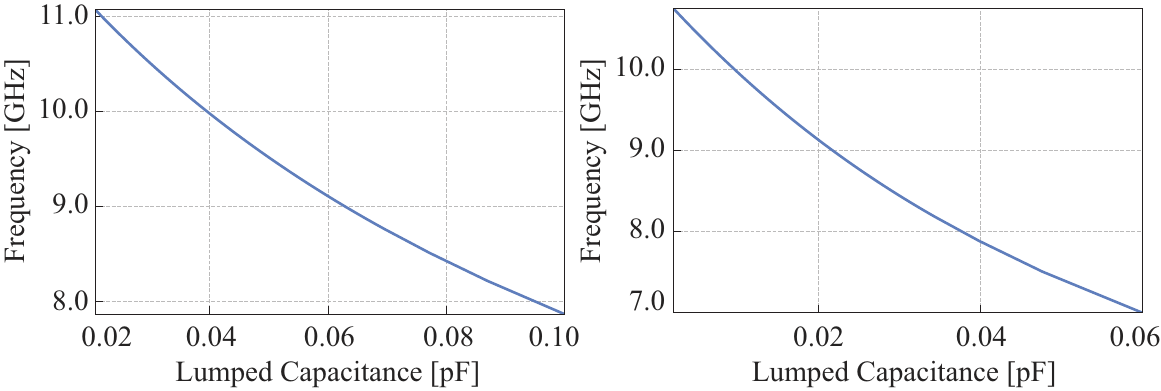}
\caption{Tuning curves for the v2 and v3 cELCs, computed by setting the denominator of \ref{eq:z_of_celc_with_cap} equal to zero.}
\label{fig:tuning_curves}
\end{figure}

From (\ref{eq:transformer_ratio}), we find $n^2=6.8$ for the v2 element and $n^2=4.05$ for the v3 element (Table \ref{table:celc_fit_pars}. The result is that the transformed lumped capacitance for the v2 element should be $C'_p=n^2 0.02=0.136$ pF, and $C'_p=n^2 0.02=0.08$ pF for the v3 element. General tuning curves of resonance frequency versus $C_p$ can be generated by solving for the roots of the denominator of (\ref{eq:z_of_celc_with_cap}). The tuning curves for the v2 and v3 elements are shown in Fig. \ref{fig:tuning_curves}, revealing a smoothly varying shift to lower frequencies as the capacitance is increased. These tuning curves illustrate part of the tradeoffs in tuning strategies, with the v3 element exhibiting a much greater tuning range for the same added capacitance. The larger range results from the larger inductance in the element. As discussed earlier, the tradeoff for changing the ratio of the element inductance to capacitance is a change in the $Q$, so that tuning range and instantaneous bandwidth are correlated properties that can be addressed during the element design.

For the added $0.02$ pF capacitor $C_p$, the resonance would be expected to shift to $9.9$ GHz for v2 and to $9.2$ GHz for v3. Despite the approximate theory developed leading to the expression for the transformer ratio (Eq. \ref{eq:transformer_ratio}), the agreement is quite good. This agreement, which persists for the various cELC designs with varying parameters, suggests that complete parametric sweeps may not be necessary in general once the circuit model (and all parasitics) have been established. In reality, exact agreement of the $n^2$ factor is not likely, but the predicted shift can be expected to be relatively close (as evidenced by these examples), possibly within the general error of simulation. To obtain a more precise estimate, two simulations can be performed and the factor $n^2$ determined from the actual frequency shift. For example, the value of $n^2$ that achieves the agreement shown in Fig. \ref{fig:cELC_tuning_examples} is $0.8$, which differs from the value of $0.68$ predicted by Eq. \ref{eq:transformer_ratio}. However, this factor used with Eq. \ref{eq:impedance_celc_with_cap} then correctly predicts the cELC design with other values of $C_p$.

\section{Conclusion}
In this work, we have extended the connection between polarizability and equivalent circuit models of small, radiating, waveguide-fed apertures to those that possess intrinsic geometric resonances. We have focused specifically on the cELC element, which has become a commonplace in waveguide-fed metamaterial designs and lends itself to an intuitive equivalent circuit model. While extensive analytical theories for aperture coupling between waveguides and resonant cavities have been previously presented, there does not appear to be similar studies on intrinisically resonant apertures; such apertures would not natural for general coupling problems and seem to be more suited for radiating structures. Our emphasis here has not been to attempt a direct calculation of the aperture polarizability or circuit parameters, but rather to infer these quantities based on simulations of the scattering parameters from cELC irises. The accuracy of the method is seen to be quite good, revealed by the excellent agreement between computed and fitted polarizability and S-parameter curves.

The reality of the derived circuit parameters may be debated, as there are many potential sources of inductance and capacitance in the element. Indeed, to apply such relatively simple models as presented here, the cELC structure must be designed so that the inductive and capacitive regions are easily identified and no other significant effective reactive elements are disregarded. Such reasoning leads to the relatively straightforward models presented here for the connected and disconnected cELCs. 

The crucial test of the circuit model---and the underlying motivation for this work---is whether the effective circuit parameters obtained for the cELC can be combined with external, integrated lumped elements, such as the capacitors used for illustration. The general agreement found for the frequency shifts and line shapes when a lumped element is inserted to the capacitive gap of a cELC element indicates the viability of the circuit model. To obtain the correct circuit model, modifications are necessary to the general theory of small apertures that enable the polarizability to be estimated for a resonant structure; these modifications contain assumptions regarding the field localization and enhancement within the resonant cELC. The impact of this theory is to introduce a transformer section to the circuit and an associated coupling factor, or turn ratio $n^2$. While we provide a rough estimate for this factor, which appears to provide useful guidance, there is nevertheless some ambiguity that is best resolved with one or two additional simulations to allow more precise determination of this factor.

Building on the previous work of the slot element \cite{smith2024circuit}, the present analysis that introduces circuit models for the cELC class of elements provides even more evidence that quite complex circuits can be combined with metamaterial irises for greater functionality that can be easily predicted and understood. While we have used simple cELC models and a very simple lumped element consisting of a capacitor and parasitic inductance, there is no reason that this work could not be extended to more complex lumped components, both active as well as passive. Once these models are constructed, the radiative properties of an aperture consisting of such elements can then be easily and efficiently computed using a coupled dipole formalism---a topic that will be pursued in future work.

\appendix[Calculation of the cELC Polarizability]

There are a number of procedures presented in the literature to compute the behavior of small apertures in conducting planes separating two regions. One general and rigorous procedure has been presented in \cite{mautz1983admittance}. In this formulation, the scattering problem in either of the two regions connected by the aperture can be solved independently, with an electric boundary condition replacing the aperture and assuming an unknown magnetic current $\vec{M}$ on one side of the aperture, and a magnetic current $-\vec{M}$ on the other side of the aperture. This magnetic current produces fields in both of the regions, thus setting up a self-consistency condition that enables the magnetic current to be determined.

Since the unknown magnetic current, defined as $\vec{M}=-\hat{z} \times \vec{E}$, is of opposite sign on either side of the aperture, there is no discontinuity of the magnetic field across the aperture. Continuity of the magnetic field across the aperture then requires that

\begin{equation}
    H_a(M) + H_b(M)=-H_i,
    \label{eq:continuity_of_magnetic_field}
\end{equation}

\noindent where all fields are assumed to lie along the $\hat{x}$ direction, since we assume the electric dipole and second magnetic dipole are insignificant. In this formulation, $H_a$ is the reaction field within the waveguide, which we refer to as region $a$, assuming an impressed magnetic current $M$ over the aperture, with the open boundary replaced by an electric boundary; likewise, $H_b$ is the reaction field due to radiation in free space, or region $b$, assuming a magnetic current $-M$ over the aperture, again with the open boundary replaced by an electric boundary. The magnetic current also drives the cELC resonator, which produces an additional reaction field within the gap. This field can be placed on either side of the aperture; for convenience, we place this reaction field on the region $b$ side.

Since the aperture is electrically small, we make the approximation that all quantities are evaluated at a point at the center of the aperture. In this case, that point is taken as the center of the cELC. Since we require the fields and magnetic current only at a single point they no longer have a spatial dependence and are simply complex numbers. The magnetic current can be expressed as $\vec{M}=M \delta(x-x_0) \delta(z-z_0) \hat{x}$. The magnetic current relates to the magnetic dipole moment according to

\begin{equation}
    M = j \omega \mu m.
\end{equation}

The general approach for solving \ref{eq:continuity_of_magnetic_field} in \cite{harrington1976, mautz1983admittance} involves discretizing the aperture and applying the method of moments to compute the spatially varying tangential magnetic current. Expanding the magnetic current in terms of a set of test functions, $W_i$, with unknown coefficients $V_i$, Eq. \ref{eq:continuity_of_magnetic_field} can be converted to a matrix equation by multiplying both sides by a test function $W_j$ and integrating over the aperture. For the small aperture considered here, we approximate the magnetic current by a single term $W_0$, so that

\begin{equation}
    M = V_0 W_0.
\end{equation}

\noindent Following \cite{liang1983generalized}. We define generalized admittances as

\begin{equation}
    Y_a = H_a W_0,
\end{equation}

\noindent and

\begin{equation}
    Y_b = H_b W_0.
\end{equation}

\noindent The right-hand side of Eq. \ref{eq:continuity_of_magnetic_field} can then be written as $W_0 H_i$. Combining the above equations leads to

\begin{equation}
    (Y_a + Y_b) M = W_0^2 H_i
    \label{eq:admittance_equation}
\end{equation}

Similar to \cite{liang1983generalized}, we select the test function to be 

\begin{equation}
    W_0=\sqrt{\frac{\alpha_{m0}}{\omega \mu}},
\end{equation}

\noindent with

\begin{equation}
    \alpha_{m0} = \frac{a b}{2 \omega \mu}
\end{equation}

\noindent The normalization chosen here differs from that of \cite{liang1983generalized} as the units of admittance in this treatment are in inverse Ohms rather than unitless.

Our goal is not to determine an analytical form for the polarizability, but rather to determine a network representation of the aperture. For this purpose, we rewrite (\ref{eq:admittance_equation}) as

\begin{equation}
    jB_a M + G_a M + jB_b M + G_b M = W_0^2 H_i.    
\end{equation}

\noindent where $Y_a=G_a+j B_a$ and $Y_b=G_b+j B_b$. The reason for writing the equation in this manner is that we can readily identify the admittances due to the reaction fields in the two regions. The conductance terms relate to the radiative reaction fields, while the susceptance terms relate to the effective circuit parameters that we can infer from the geometry.

A modification that must be inserted into the above equation is specific to metamaterial elements (or irises) that support intrinsic resonances. A cELC in a conducting plane between two infinite half-spaces supports a resonance, at which the electric field within the gap of the cELC will be strongly enhanced relative to the element off-resonance. As with cavity resonators \cite[Chapter~7]{collin1990field} there will be a reaction field associated with the resonator acting at the position of the magnetic current. While the form of the impedance of the resonator can be found simply from $C_i$ and $L_i$, the coupling of the current element to the resonator is not known and requires the insertion of a coupling constant $n^2$. While we do not provide a complete calculation of this coupling constant here, we expect it to be related to the enhanced field within the gap when the cELC is resonant. The continuity of the magnetic field then becomes

\begin{equation}
    jB_a M + G_a M + j n^2 B_b M + G_b M = W_0^2 H_i.
\end{equation}

The radiation damping terms can be found straightforwardly as

\begin{equation}
    G_a = \frac{1}{R_a} = \frac{\beta}{2 \mu \omega},
\end{equation}

\noindent and

\begin{equation}
    G_b = \frac{1}{R_b} = \frac{a b k^3}{6 \pi \mu \omega}.
    \label{eq:radiation_conductivity}
\end{equation}

\noindent The susceptance on the waveguide side of the aperture can be approximated primarily as inductive, so that

\begin{equation}
    B_a = \frac{1}{X_a} = \frac{1}{-\omega L_e}.
\end{equation}

\noindent Within the cELC, the admittance relates to the series $LC$ combination, or

\begin{equation}
    B_b = \frac{1}{X_b}=\frac{1}{-\omega L_i + 1/(\omega C_i)}.
\end{equation}

\noindent Combining the above equations leads to the expression for the polarizability in Eq. \ref{eq:celc_polarizability}.

The final step is to determine the factor $n^2$, which indicates the field enhancement at the resonance frequency of the cELC. The quality factor of the resonance can be expressed as 

\begin{equation}
    Q = \omega_0 \frac{\varepsilon_0\int{ E^2 dV}}{P_{\textrm{rad}}}.
\end{equation}

\noindent As described in the text above, at resonance, the power radiated is fixed for a given incident waveguide field, regardless of the quality factor. This constancy of the radiated power term indicates we can compare the field enhancement for different $Q$ values. The ratio of Q factors for two different resonators is

\begin{equation}
    \frac{Q_{\textrm{res}}}{Q_0} = \frac{\int{ E_r^2 dV}}{\int{ E_0^2 dV}}.
\end{equation}

\noindent Assuming the field distributions are concentrated in the aperture plane and are essentially uniform (specifically across the capacitive gap), we arrive at

\begin{equation}
    \frac{E_r}{E_0} = \sqrt{\frac{Q_{\textrm{res}}}{Q_0}} = n^2.
\end{equation}

\noindent If the quality factor of one of the cELC designs can be approximated as $Q_0 \approx 1$, then we obtain the result

\begin{equation}
    n^2 \approx \sqrt{Q_{\textrm{res}}} = \sqrt{\omega_0 R_{\textrm{rad}}C} = \sqrt{\omega_0 \frac{C}{G_{\textrm{rad}}}}. 
\end{equation}

\noindent Using the expression for radiation resistance, Eq. \ref{eq:radiation_conductivity}, we find

\begin{equation}
    n^2 \approx \sqrt{ \frac{6 \pi}{a b k} \frac{C}{\varepsilon}}.
    \label{eq:transformer_ratio}
\end{equation}

\noindent Note that for two different cELC elements, the ratio of their coupling factors can be found to good approximation as

\begin{equation}
    \frac{n_1^2}{n_2^2}=\sqrt{\frac{\omega_2 C_1}{\omega_1 C_2}}.
\end{equation}

To more accurately assess the coupling factor, a cavity coupling approach such as that described in \cite[Chapter~5]{collin1990field} could be used. The inferred dependence of the cELC mode on Q is somewhat universal for cavity coupling by small apertures, though a more thorough treatment in the case of cELCs is warranted.

\section*{Acknowledgment}

This work was partially funded by Kymeta Corporation.

\bibliographystyle{IEEEtran}
\bibliography{bibtex/bib/IEEEexample}

\end{document}